\documentclass[prd,aps,twocolumn,nofootinbib,preprintnumbers,superscriptaddress,preprintnumbers,balancelastpage,longbibliography]{revtex4-2}
\usepackage[english]{babel}

\usepackage{amsmath,mathtools,physics,xfrac}
\usepackage{graphicx}
\usepackage{afterpage}
\usepackage{float}
\usepackage{subfigure}
\usepackage{rotating}
\usepackage{multirow}
\usepackage{tabularx}
\usepackage{booktabs}
\usepackage{fancyhdr}
\usepackage[utf8]{inputenc}
\usepackage{theorem}
\usepackage{moreverb}
\usepackage{euscript}
\usepackage{psfrag}
\usepackage{slashed}
\usepackage{mathtools}
\usepackage{makecell}
\usepackage{adjustbox}
\usepackage{dcolumn}
\usepackage{bm}
\usepackage[dvipsnames]{xcolor}
\usepackage{graphics}
\usepackage{hyperref}
\usepackage{lettrine}
\usepackage{aas_macros}

\hypersetup{
     colorlinks   = true,
     citecolor    = orange,
     urlcolor     =  orange,
     linkcolor    =  orange
}

\usepackage{color,xcolor}

\definecolor{darkblue}{rgb}{0.0,0.0,0.75}
\definecolor{darkred}{rgb}{0.6,0.0,0}
\definecolor{darkgreen}{rgb}{0.0,0.6,0.}
\definecolor{nice}{rgb}{0.8,0, 0.8}

\newcommand{\hp}[0]{$H^+_3\;$}

\input Zallman.fd

\LettrineTextFont{\itshape}
\begin{document}

\preprint{SLAC-PUB-17753}

\title{Search for Dark Matter Ionization on the Night Side of Jupiter with Cassini}

\author{Carlos Blanco}
\thanks{\href{mailto:carlosblanco2718@princeton.edu}{carlosblanco2718@princeton.edu}, \href{http://orcid.org/0000-0001-8971-834X}{0000-0001-8971-834X}}
\affiliation{Princeton University, Department of Physics, Princeton, NJ 08544, USA}
\affiliation{Stockholm University and The Oskar Klein Centre for Cosmoparticle Physics, Alba Nova, 10691 Stockholm, Sweden}

\author{Rebecca~K.~Leane}
\thanks{\href{mailto:rleane@slac.stanford.edu}{rleane@slac.stanford.edu}, \href{http://orcid.org/0000-0002-1287-8780}{0000-0002-1287-8780}}
\affiliation{Particle Theory Group, SLAC National Accelerator Laboratory, Stanford, CA 94035, USA}
\affiliation{Kavli Institute for Particle Astrophysics and Cosmology, Stanford University, Stanford, CA 94035, USA}

\date{\today}

\begin{abstract}
We present a new search for dark matter using planetary atmospheres. We point out that annihilating dark matter in planets can produce ionizing radiation, which can lead to excess production of ionospheric $H_3^+$. We apply this search strategy to the night side of Jupiter near the equator. The night side has zero solar irradiation, and low latitudes are sufficiently far from ionizing auroras, leading to an effectively background-free search. We use Cassini data on ionospheric $H_3^+$ emission collected 3 hours either side of Jovian midnight, during its flyby in 2000, and set novel constraints on the dark matter-nucleon scattering cross section down to about $10^{-38}$~cm$^2$. We also highlight that dark matter atmospheric ionization may be detected in Jovian exoplanets using future high-precision measurements of planetary spectra.
\end{abstract}

\maketitle

\lettrine{E}{arly} on an autumn morning of 1997, the Cassini spacecraft launched from Cape Canaveral aboard the Titan IV-B rocket, beginning the seven-year journey to Saturn and its majestic icy rings. To get there, Cassini was powered by the heat from nuclear decays of onboard Plutonium-238. But that wasn't all; gravitational slingshot assists were also exploited from Venus (twice), Earth, and Jupiter. Equipped with instruments to record data from radio waves to the extreme ultraviolet (EUV), Cassini would capitalize on these flybys to extract multi-wavelength data on multiple solar system bodies~\cite{2004Icar..172....1H,2001JGR...10630099B}.

Using Cassini's Visual and Infrared Mapping Spectrometer (VIMS), one such measurement was levels of trihydrogen cations, known as $H_3^+$. These ions are highly abundant throughout the Universe, and are produced from $H_2$ interactions with cosmic rays, EUV stellar irradiation, planetary lightning, or electrons accelerated in planetary magnetic fields~\cite{RevModPhys.92.035003}. Planetary $H_3^+$ levels have been extensively studied (see $e.g.$ Refs.~\cite{1989Natur.340..539D,1993ApJ...405..761T,1993ApJ...408L.109G,1999ApJ...524.1059T,1999ApJ...521L.149S,2015JGRA..120.6948S,2013EPSC....8..939B, 2002EGSGA..27.2120E,2011epsc.conf..824M,2011JGRA..116.9306G, 2015Icar..245..355M, 2017GeoRL..4411762O,RevModPhys.92.035003}), and they are important as they provide vital insights into atmospheric temperature, as well as a tracer of electric currents running through the atmosphere~\cite{RevModPhys.92.035003}. 

In this work, we point out that dark matter can produce an additional source of $H_3^+$ in planetary atmospheres. This will be produced if dark matter scatters and is captured by planets, and consequently annihilates, producing ionizing radiation. To produce detectable $H_3^+$, the dark matter must produce ionization in the planet's ionosphere, which occurs for the part of the captured dark matter distribution already thermalized towards the surface~\cite{Leane:2022hkk}. Alternatively, dark matter may also annihilate to mediators with lifetimes or kinematic boosts that lead to frequent decays away from where dark matter is thermalized~\cite{Batell:2009zp,Pospelov:2007mp,Pospelov:2008jd,Rothstein:2009pm,Chen:2009ab,Schuster:2009au,Schuster:2009fc,Bell_2011,Feng:2015hja,Kouvaris:2010,Feng:2016ijc,Allahverdi:2016fvl,Leane:2017vag,Arina:2017sng,Albert:2018jwh, Albert:2018vcq,Nisa:2019mpb,Niblaeus:2019gjk,Cuoco:2019mlb,Serini:2020yhb,Mazziotta:2020foa, Leane:2021ihh, Bell:2021pyy,Leane:2021tjj,Li:2022wix,Acevedo:2023xnu}, such that the sum of these scenarios covers a wide range of the parameter space.

\begin{figure}[t!]
    \centering
    \includegraphics[width=0.75\columnwidth]{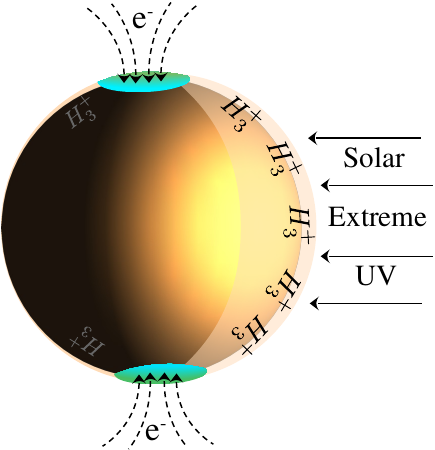}
    \caption{Schematic of \hp production in Jupiter. Auroral \hp emission near the magnetic poles is sourced by precipitating electrons, and solar extreme UV irradiates the day side and dominates \hp production near the equator. No \hp is expected from known processes at low latitudes on the night side, making it an ideal dark matter signal region.}
    \vspace{-5mm}
    \label{fig:setup}
\end{figure}

We execute our dark matter ionization search using Cassini's VIMS flyby data of Jovian ionospheric $H_3^+$~\cite{2015JGRA..120.6948S}. We target Jupiter as it is the most efficient dark matter captor compared to Saturn or other planets, and its relatively cool core allows the lightest dark matter particles to be retained~\cite{Leane:2020wob}. To optimize for signal over background, we study Cassini data taken 3 hours either side of Jovian midnight, which eliminates the solar EUV irradiation background. We also focus on low-latitude data, as latitudes near the poles are subject to the intense Jovian magnetic fields, which produce ionizing auroras which are a significant source of $H_3^+$. At low latitudes near Jovian midnight, the background expectation is effectively zero, because the recombination time of $H_3^+$ produced by solar irradiation on the Jovian day side or the auroras near the poles is as fast as about $10-15$ minutes~\cite{1998JGR...10320089A}. This leaves insufficient time for $H_3^+$ to migrate from the day side to the night side, or to migrate far from the magnetic poles. These effects are schematically summarized in Fig.~\ref{fig:setup}.

We will also investigate the future potential to discover dark matter using atmospheric ionization in Jovian exoplanets. This is most advantageous in the inner Galaxy, where the dark matter density is expected to be higher. However, it is impossible to spatially resolve the low-latitude emission in this case, and so auroral emission will dominate as the ionization background. We will show that the advantage of higher dark matter densities, as well as using larger exoplanets such as super-Jupiters which more readily capture dark matter, can overcome the significant auroral backgrounds. This can lead to potential future dark matter discovery space down to small scattering cross sections.

Our paper is organized as follows. We first discuss how \hp is generally produced and detected through spectroscopic analyses, before discussing existing Jovian \hp measurements and expectations without dark matter. We then detail how these ionization rates are calculated for Jupiter's ionosphere, and calculate the maximum ionizing power consistent with Cassini night-side observations. We then discuss the pathways for dark matter-induced $H_3^+$ production, and explore the relevant dark matter-nucleon scattering parameter space, setting new constraints based on the requirement that the dark matter-induced $H_3^+$ is larger than Cassini's measurements. We conclude with a discussion of future opportunities to discover dark matter using planetary or sub-stellar atmospheric ionization.\\

\noindent\textbf{\textit{Astrophysical \hp Production and Emission--}} The dominant species in the high-altitude Jovian atmosphere is neutral molecular $H_2$. The creation of \hp in the Jovian ionosphere follows from interactions of high-energy particles, ions, or ionizing photons with $H_2$~\cite{RevModPhys.92.035003}. While there are a few \hp production pathways with different ionization sources, an example of ionizing scattering is 
\begin{equation}
    H_2 + e^{-*} \rightarrow H_2^+ + 2e^-,
\end{equation}
where high-energy electrons $e$ can be injected by e.g. electrons accelerated in Jovian magnetic fields, or the annihilation of dark matter.  In any case, once $H_2^+$ is produced, it reacts almost instantaneously with the surrounding $H_2$, creating \hp via \cite{RevModPhys.92.035003,wu2019status}
\begin{equation}
    H_2^+ + H_2\rightarrow H_3^+ + H.
\end{equation}
This \hp then exists in the planetary ionosphere above a layer called the homopause, where Jovian gas decouples from convective mixing currents. This is because the relative abundance of gases above the homopause is free to evolve with altitude due to the lack of mixing, and so heavier gases that can destroy \hp (such as methane) rapidly decrease in concentration~\cite{sanchez2010introduction}. Below the homopause, \hp is quickly destroyed by protonating other molecules, and therefore is not relevant~\cite{strobel_atreya_1983,1998JGR...10320089A}.

Once \hp has been produced, it is expected to approach a quasi-thermal equilibrium, wherein the radiative deexcitation of vibrational states is slow compared to their population by collisional excitation with the surrounding $H_2$~\cite{miller1990infrared,2015JGRA..120.6948S}. The $H_2$ has characteristic temperatures between about $700^\circ$~K and $1000^\circ$~K in the Jovian ionosphere~\cite{2004jpsm.book..185Y}, which is inherited by the $H_3^+$ before emitting infrared radiation efficiently through radiative vibrational deexcitation. The infrared spectrum of \hp contains a large amount of substructure, and its emission lines are strongest in the spectral window between about 3~$\mu$m and 5~$\mu$m. Since this is also the spectral region of peak thermal emission for a black body of temperature between about $700^\circ$~K and $1000^\circ$~K, \hp acts as an extremely efficient thermal radiator around these temperatures~\cite{RevModPhys.92.035003,miller2006driver}.

Spectroscopic analysis of \hp relies heavily on the \textit{relative} intensities of the emission lines, since theoretical modeling is subject to significant uncertainty in the \textit{intrinsic} intensity of each individual peak, although their frequencies and widths can be modeled to an exquisite degree of precision~\cite{RevModPhys.92.035003,2015JGRA..120.6948S}. Specifically, the relative intensities of the infrared emission lines can be used to determine the temperature of the gas, assuming local thermodynamic equilibrium. In what follows, the temperatures used for \hp have been calculated from the measured relative intensities of the lines.

\hp plays an important role in the thermal balance of gas planets like Jupiter and Uranus, where its emission acts as the thermostatic mechanism in the high-altitude atmosphere~\cite{RevModPhys.92.035003,miller2006driver}. While this energy equilibrium is borne of a complicated dynamical atmospheric system, empirically the net effect of additional ionizing energy injection is an increased \hp emission. Therefore, we will adopt the simplified but empirical model of energy balance where the total surface infrared emission of \hp is proportional to the ionizing energy injected into the atmosphere~\cite{2000Icar..147..366R}.
\\

\noindent\textbf{\textit{Ionization Data and Constraints from Cassini--}} 
The production of \hp due to ionizing radiation was first confirmed spectroscopically in the auroral ionosphere of Jupiter, using the Voyager Ultraviolet Spectrometer experiment~\cite{1989Natur.340..539D}. While the strongest \hp emission (therefore greatest production) is found near the poles, in the auroral regions which are powered by infalling ions, there is a detectable smooth \hp infrared signal found extending up to the equator on the day side~\cite{2000Icar..147..366R}. Volcanic eruptions from Jupiter's third-largest moon Io also deposit ionizing radiation into Jupiter's atmosphere, however this is a characterized local phenomenon.

The equatorward ($\pm20^\circ$ latitude) \hp emission signal is produced from \hp originating from solar EUV~\cite{2000Icar..147..366R,1998JGR...10320089A}, which has an ionizing power of $P_{\text{ion}}^\text{EUV} = 62 \; \mu \text{W/m}^2$ for wavelengths between Lyman-$\alpha$ and 10 nm. Since the dissociative recombination time of \hp in the upper Jovian atmosphere is much shorter ($\sim10-15$ minutes at the altitude of peak \hp emission) than the half-day period~\cite{1998JGR...10320089A}, it is expected that effectively all of the \hp produced during the day will disappear during the night, and there is essentially no auroral \hp contamination at low latitudes. Therefore, there are no known ionization sources expected at Jovian night time at low latitudes, allowing for a striking potential dark matter-induced \hp signal.

Cassini is an ideal instrument to probe a potential dark matter-induced \hp signal, as its flyby collected data on equatorward \hp on the night side of Jupiter. In this target region, Cassini did not detect any \hp\cite{2015JGRA..120.6948S}, consistent with Standard Model expectations. Therefore, we can place constraints on the homogenous ionization of the night side hydrogen based on the sensitivity of Cassini's observation, which is taken to be the uncertainty of the reported equatorward measured (no detection) intensity,
\begin{equation}
\label{eq:Imax}
    I^{\text{H}_3^+} < 0.03 \,  \text{mW}/\text{m}^2/\mu \text{m}/\text{sr}.
\end{equation}
To set constraints using the night-side observations of the Jovian atmosphere, we predict what the expected \hp emission signal would be as a function of the injected ionizing power. We adopt a model of the total surface \hp emission $E^{\text{H}_3^+}(T)$ in which the number of \hp molecules produced (and therefore the column density $CD^{\text{H}_3^+}$) is proportional to the ionizing power $P_{\text{ion}}$ and the temperature-dependent molecular emission $E_{\text{mol}}(T)$, which is calculated using the cooling function derived by Ref. \cite{miller2013cooling}. The total surface emission is given by
\begin{equation}
    E^{\text{H}_3^+}(T) \propto CD^{\text{H}_3^+}(P_{\text{ion}}) \times E_{\text{mol}}(T),
\end{equation} 
with
\begin{equation}
   \text{ln}(E_{\text{mol}}(T)) = \sum_{n} c_n T^n,
\end{equation} 
where $c_n$ are fitting coefficients. This parametrization of the cooling function is known to be accurate to less than $0.5\%$ for temperatures between 300~K and 1800~K~\cite{miller2013cooling}. Since the measured emission $I^{\text{H}_3^+}$ of \hp is linearly related to the total surface emission up to geometric factors, we can express the measurable intensity as a function of temperature and ionizing power as
\begin{equation}
\label{eq:alphabeta}
    I^{\text{H}_3^+} = \alpha \times \beta \times P_{\text{ion}} E_{\text{mol}}(T),
\end{equation}
where $\alpha$ and $\beta$ are linear coefficients that respectively account for the geometric factors and \hp production efficiency mentioned above. Since $\alpha$ and $\beta$ are independent of night-side or day-side conditions, they cancel when taking ratios of the day-side and night-side emission in Eq.~(\ref{eq:alphabeta}) to compute $P_{\text{ion}}^{\text{night}}$. This allows us to forego directly computing $\alpha$ and $\beta$ and also makes our results independent of the associated systematic uncertainties. We therefore use the day side emission data, along with night-side temperature measurements and Cassini's non-detection of \hp emission, to calculate the night time constraints on ionizing power,
\begin{equation}
    P_{\text{ion}}^{\text{night}} < \frac{I^{\text{H}_3^+}_{\text{max}}}{I^{\text{H}_3^+}_{\text{day}}} \times \frac{E_{\text{mol}}(T_{\text{day}})}{E_{\text{mol}}(T_{\text{night}})} \times P_{\text{ion}}^\text{EUV} \times 1.5, 
\end{equation}
where $I^{\text{H}_3^+}_{\text{max}}$ saturates the bound in Eq.~(\ref{eq:Imax}), and $I^{\text{H}_3^+}_{\text{day}} = 0.09 \; \text{mW/m}^2/\mu \text{m/sr}$ is the equatorward daytime intensity measured using ground-based observations by the Infrared Telescope Facility (IRTF)~\cite{2000Icar..147..366R}, but corrected in order to be directly compared to night-side measurements by Cassini~\cite{2015JGRA..120.6948S}. The factor of 1.5 accounts for the fact that the Jovian ionospheric model (JIM) finds that photons are about 1.5 times more efficient at producing \hp as electrons~\cite{2000Icar..147..366R,1998JGR...10320089A}. We adopt a day time (night time) equatorward temperature of $850^\circ\text{K} \pm 50^\circ\text{K}$ ($800^\circ\text{K} \pm 50^\circ\text{K}$) consistent with the predictions from day-side and night-side observations~\cite{2015JGRA..120.6948S,MILLER199757}.

We therefore find the maximum night-side ionizing power to be 
\begin{equation}
\label{eq:Pion}
    P_{\text{ion}}^{\text{night}} < 40 \pm 17 \; \mu \text{W/m}^2.
\end{equation}
This quantity should be understood to be the maximum amount of \textit{additional} ionizing power, contributing to the energy budget of the ionosphere, that is consistent with Cassini data. The uncertainty in Eq.~(\ref{eq:Pion}) is dominated by the uncertainties in the temperature measurements. Since this calculation has been totally independent of any dark matter considerations, this is the upper bound of ionizing power by any additional source. We can now compare this maximum allowed ionizing power, as set by Cassini on Jupiter's night side, to the predictions of ionizing power from annihilating dark matter.\\

\noindent\textbf{\textit{Dark Matter Induced \hp Production--}}
For dark matter to produce $H_3^+$, it must deposit its ionizing energy into the planet's ionosphere. This can be achieved in two ways. One scenario is that when dark matter is captured, it settles into an equilibrium distribution within Jupiter, with a density distribution that is usually peaked towards the core, but with a distribution tail that can still be substantial in the ionosphere~\cite{Leane:2022hkk}. The dark matter, which in equilibrium already sits in the ionosphere, can annihilate there into short-lived or not-very kinetically-boosted mediators, which immediately produce ionizing radiation, and therefore $H_3^+$. An alternative scenario is that dark matter may annihilate into longer-lived or boosted mediators, which may decay and deposit the ionizing energy at some distance from where the dark matter annihilated.

The dark matter ionizing power $P_{\rm ion}^{\rm DM}$ can be compared to the maximum night-side ionization in Eq.~(\ref{eq:Pion}) to obtain a limit on the DM-induced ionization. The dark matter ionizing power is calculated as
\begin{equation}
\label{eq:PionDM}
    P_{\rm ion}^{\rm DM}=\frac{\Gamma_{\rm ann}\times f_{\rm iono}}{4\pi R^2},
\end{equation}
where $\Gamma_{\rm ann}$ is the total equilibrium mass annihilation rate in the planetary volume, $f_{\rm iono}$ is the fraction of annihilation events occurring in the ionosphere of the planet, and $R$ is the planet radius.  For the case where the annihilation occurs in the core, and the annihilation products are boosted to the ionosphere, or the mediator is sufficiently long-lived to decay in the ionosphere, $f_{\rm iono}\sim1$ is easily achieved across a wide range of parameters. For the case of short-lived mediators, or mediators with small kinematic boosts, $f_{\rm iono}$ is generally less than one, and results can be rescaled accordingly~\cite{Leane:2022hkk}.

The equilibrium mass annihilation rate is obtained when dark matter annihilation and capture are in equilibrium. This is given by~\cite{Leane:2020wob}
\begin{equation}
    \Gamma_{\rm ann}= f_{\rm cap}\times\pi R^2 \rho_\chi v_\chi \sqrt{\frac{8}{3 \pi}} \left( 1 + \frac{3}{2} \frac{v_{\rm esc}^2}{v_\chi^2}  \right) \, ,
\end{equation}
where $m_\chi$ is the dark matter mass, $\rho_\chi=0.4$~GeV/cm$^3$ is the local dark matter density, $v_\chi=270$~km/s is the local dark matter velocity dispersion, $v_{\rm esc}$ is the planetary escape velocity, and $f_{\rm cap}$ is the fraction of particles passing through the planet that are captured. For the maximal geometric rate, corresponding to sufficiently large dark matter-nucleon scattering cross sections, $f_{\rm cap}\approx1$. As the scattering cross section decreases, $f_{\rm cap}$ also decreases. We calculate the captured fraction of dark matter particles that are above our Cassini data's detection threshold as discussed in the previous section, and set limits by linking it to the dark matter nucleon scattering cross section using the \texttt{Asteria} package~\cite{Leane:2023woh}.\\

\begin{figure}[t!]
    \centering
    \includegraphics[width=\columnwidth]{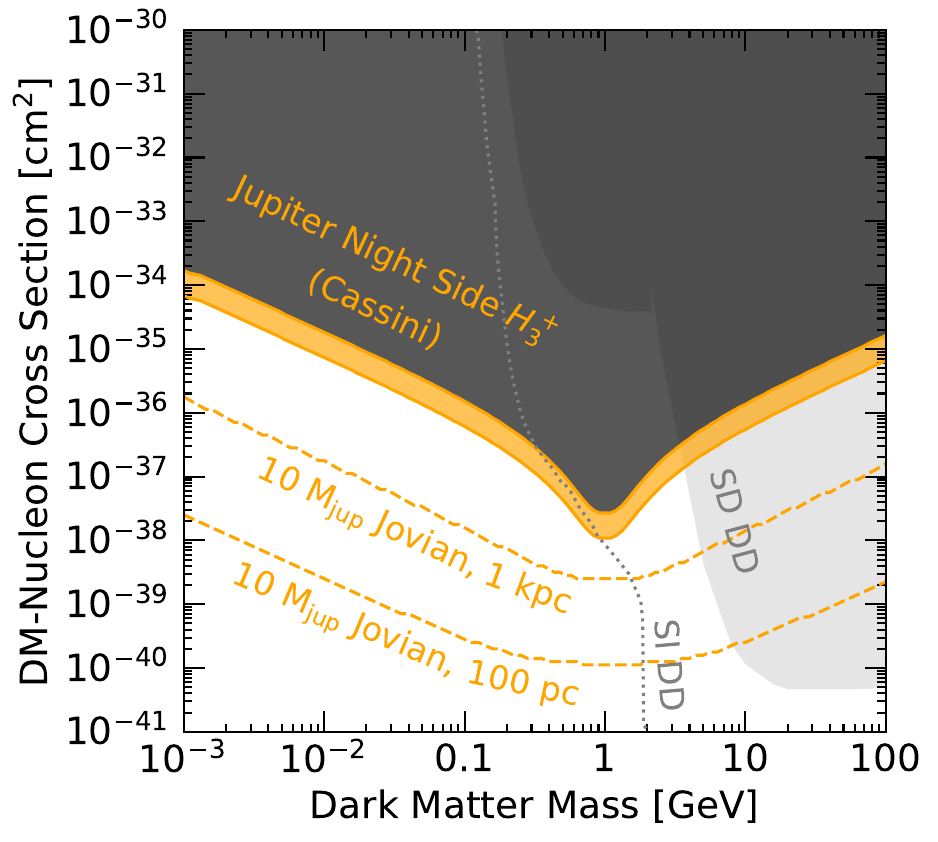}
    \caption{New constraints on the dark matter-nucleon scattering cross section using planetary ionosphere measurements. The orange band covers our uncertainty in limits with night side Cassini VIMS Jovian $H_3^+$ measurements; the dark-shaded region is excluded. We also show sensitivity projections for a benchmark Jovian exoplanet at 1 kpc or 100 pc from the Galactic center (dashed), see text for details. We also show limits from direct detection, spin-independent ``SI DD" and spin-dependent proton scattering ``SD DD".}
    \label{fig:est}
\end{figure}

\noindent\textbf{\textit{Dark Matter Parameter Space--}}
Figure~\ref{fig:est} shows our calculated limits on the dark matter-nucleon scattering cross section as a function of dark matter mass, using our ionization search strategy. The dark-shaded region corresponds to the Jovian night side limits we derive from Cassini VIMS flyby data, labeled ``Jupiter Night Side $H_3^+$ (Cassini)", with the orange band covering the uncertainty on this limit. As we will discuss, the exact shape of the bounds in Fig.~\ref{fig:est}  can vary with particle physics models. Despite this, the search strategy that we present here can be many orders of magnitude more sensitive than existing searches, especially for light dark matter. For example, in this figure, we have taken $f_{\rm iono}\sim1$, which corresponds to the dark matter-rest mass energy being deposited fully in the ionosphere, but this can be easily rescaled depending on the dark matter model of interest. For example, in the case of a one-mediator model which is short-lived and also not kinematically very boosted, $f_{\rm iono}\ll1$ and the bounds would apply only at larger cross sections. We also assume dark matter annihilation into electrons, though annihilation into any ionizing species will be efficient. These bounds will be largely unchanged if the final state is instead hadronic.

In Fig.~\ref{fig:est} we also show projected sensitivities to the dark matter-nucleon scattering cross section for a dark matter ionization search, for benchmark inner-Galaxy Jovian exoplanets, $i.e.$ super-Jupiters with 10 times Jupiter's mass. For these example sensitivities we have simply assumed that the dark matter signal matches or exceeds the ionization background. Cosmic-ray ionization is not expected to exceed 100 times the value measured near the solar position~\cite{peron2021probing,huang2021gev}. We therefore assume that the Jovian auroras dominate the background, with the auroral ionization intensity of these benchmarks being the same as Jupiter's, which is about a factor of 100 higher than the solar EUV~\cite{1986aiop.book.....A}.  This is required as the low latitudes exploited for our Cassini Jupiter search cannot be disentangled from the auroral emission at large distances. While the inner-Galaxy Jovian exoplanets have larger ionization backgrounds, they are also embedded in larger dark matter densities than Jupiter, leading to larger sensitivities. Furthermore, super-Jupiters are more massive than Jupiter and have the additional bonus of larger capture rates, leading to larger expected dark matter signals. Our benchmark exoplanets are shown at 100 pc and 1 kpc, and in the optimistic case that further towards the Galactic center is detectable, we see that the dark matter parameter space sensitivity can substantially increase. For the inner Galaxy dark matter distribution, we have assumed a standard NFW dark matter profile.

Fig.~\ref{fig:est} assumes no evaporation, as the minimum dark matter mass that can be retained without evaporating is model dependent~\cite{Acevedo:2023owd}. Given the focus of our work is pointing out this new search strategy, we do not focus on any detailed particle models, but provide some characteristic numbers in some benchmark models for context. For contact-interaction dark matter models, in our parameter range shown, the evaporation mass ranges between about $\sim200$ MeV$-1$ GeV. For long-range attractive interaction models, the dark matter evaporation mass can be instead sub-MeV~\cite{Acevedo:2023owd}. It is also important to note that we conservatively assumed capture rates in the contact-interaction scenario; including additional captured dark matter from an attractive long-range model can boost our rates considerably by multiple orders of magnitude, depending on the specific model parameters.

There are other existing search strategies that complementarily probe our parameter space in Fig.~\ref{fig:est}. The strongest Earth-based bounds are from direct detection experiments. We show the latest strongest existing bounds from spin-dependent proton-dark matter scattering, which are from CRESST-III~\cite{CRESST:2022dtl} and PICO-60~\cite{PICO:2019vsc}, as well as spin-independent nucleon-dark matter scattering bounds, from CRESST~\cite{CRESST:2019jnq,CRESST:2022lqw}, DarkSide~\cite{DarkSide:2018bpj}, XENON-nT~\cite{XENON:2023cxc}, and LZ~\cite{LZ:2022ufs}.

Other complementary astrophysical constraints include celestial-body searches with gamma rays, including with Jupiter~\cite{Leane:2021tjj}, brown dwarfs~\cite{Leane:2021ihh}, and white dwarfs~\cite{Acevedo:2023xnu}. These works derive limits assuming annihilation directly into gamma rays. Our search is instead most sensitive to annihilation into massive ionizing species such as electrons and protons, which have the best ionizing power. Models where the final state gamma-ray spectrum becomes flattened will generally be less constrained using gamma-ray observations. Since we simply require the spectrum to be above the ionizing threshold, which is orders of magnitude smaller than the dark matter mass, our approach serves as a complementary way to probe the dark matter annihilation. We are also only sensitive to the much more precisely known local density of dark matter, in contrast to inner Galaxy searches~\cite{Leane:2021ihh,Acevedo:2023xnu,John:2023knt}. Different dark matter profile assumptions can wipe out the sensitivity of such searches, but would not affect our night-side Jovian search with Cassini. Additionally, this work is complementary to constraints set from electrons detected in Jupiter's radiation belt, since our results do not depend on the details of the time-varying magnetic field around Jupiter~\cite{Li:2022wix}. See also $e.g.$ Ref.~\cite{YanLiFan} for complementary dark photon constraints.\\

\noindent\textbf{\textit{Conclusions and Future Opportunities--}}
We have pointed out and shown for the first time that dark matter can produce ionizing radiation in planetary atmospheres, which is detectable through a smoking-gun excess of atmospheric trihydrogen cations, $H_3^+$. We used Cassini data to search for our new dark matter signature, exploiting flyby low-latitude $H_3^+$ data of Jupiter's night side, which is effectively background free due to the lack of solar irradiation, as well as the low-latitudes being resolved sufficiently far away from the highly-ionizing Jovian auroras near the poles. Night-side Cassini data provides new constraints on the dark matter-nucleon scattering cross section, unprobed by any other experiment. Improved Jovian measurements compared to Cassini are planned in the 2030s with the European Space Agency's Jupiter Icy Moons Explorer (JUICE)~\cite{2023SSRv..219...53F}, which may allow increased sensitivity to dark matter ionization and consequently the dark matter parameter space in future.

We also showed that exoplanets in more dense dark matter environments such as the inner Galaxy could provide even stronger sensitivities to dark matter atmospheric ionization. While an inner Galaxy search would not benefit from low-latitude searches, as auroras would not be able to be distinguished with the resolution of current telescopes, the dark matter ionization rate still can be significantly larger than auroral backgrounds. This is also aided by larger dark matter capture rates being possible with larger planets than Jupiter, such as super Jupiters. Although optimistic, a future spectral measurement of exoplanets in the inner Galaxy with $e.g.$ JWST, the Roman Space Telescope, or a future more sensitive telescope could realize the exoplanetary sensitivities presented in Fig.~\ref{fig:est}.

Considering local exoplanets, $H_3^+$ is currently broadly searched for in the atmospheres of giant exoplanets, and is part of the target list for the European Space Agency's upcoming exoplanet mission, called the Atmospheric Remote-sensing Infrared Exoplanet Large-survey (ARIEL)~\cite{2018ExA....46..135T}. ARIEL will have unprecedented sensitivity to the atmospheric chemistry of transiting gas giants, and is planned to launch in 2029. Other telescopes such as JWST, or future thirty-meter class telescopes can also detect atmospheric \hp through high-resolution spectroscopy~\cite{2022AJ....164...63G}. We expect that these telescopes and missions may therefore provide excellent sensitivity to dark matter atmospheric ionization in nearby exoplanets.

Going forward to detect dark matter atmospheric ionization, it will also be fruitful to consider other sub-stellar systems with low irradiance due to no host star, such as brown dwarfs, or gas giants on wider orbits, or gas giants that are orbiting cooler stars. Brown dwarfs have in fact been shown to have much lower auroral emission than anticipated~\cite{2018ApJ...859...74S,2022AJ....164...63G}, making them an interesting additional low-background dark matter ionization target. We will present results on detecting dark matter ionization in such sub-stellar objects outside our solar system in an upcoming publication~\cite{BlancoLeane}.\\

\noindent\textbf{\textit{Acknowledgments--}} We thank JiJi Fan and Bruce Macintosh for helpful comments and discussions. This work was initiated at the Aspen Center for Physics, which is supported by National Science Foundation grant PHY-1607611. R.K.L. is supported by the U.S. Department of Energy under Contract DE-AC02-76SF00515. The work of C.B.~was supported in part by NASA through the NASA Hubble Fellowship Program grant HST-HF2-51451.001-A awarded by the Space Telescope Science Institute, which is operated by the Association of Universities for Research in Astronomy, Inc., for NASA, under contract NAS5-26555.

\bibliography{ionize}
\bibliographystyle{apsrev4-2}

\end{document}